\documentclass[aps,pre,twocolumn,superscriptaddress,showpacs]{revtex4}

\usepackage{graphicx}% Include figure files
\usepackage{dcolumn}% Align table columns on decimal point

\begin{document}

%\draft

\title{Scaling of non-Markovian Monte Carlo wave-function methods}

\author{J. Piilo}

\affiliation{School of Pure and Applied Physics, University of KwaZulu-Natal,
Durban 4041, South Africa}

\author{S. Maniscalco}

\affiliation{School of Pure and Applied Physics, University of KwaZulu-Natal,
Durban 4041, South Africa}

\affiliation{INFM, MIUR and Dipartimento di Scienze Fisiche ed
Astronomiche dell'Universit\`{a} di Palermo, via Archirafi 36,
90123 Palermo, Italy.}

\author{A. Messina}
\affiliation{INFM, MIUR and Dipartimento di Scienze Fisiche ed
Astronomiche dell'Universit\`{a} di Palermo, via Archirafi 36,
90123 Palermo, Italy.}

\author{F. Petruccione}
\affiliation{School of Pure and Applied Physics, University of KwaZulu-Natal,
Durban 4041, South Africa}

\date{\today}

\begin{abstract}
We demonstrate a scaling method
for non-Markovian Monte Carlo wave-function
simulations used to study open quantum systems
weakly coupled to their environments.
We derive a scaling equation,
from which the result
for the expectation values of arbitrary
operators of interest can be calculated,
all the quantities in the equation
being easily obtainable from the scaled
Monte Carlo wave-function simulations.
In the optimal case, the scaling method can be used,
within the weak coupling approximation,
to reduce the size of the generated
Monte Carlo ensemble
by several orders of magnitude. Thus,
the developed method allows faster simulations
and  makes it possible to solve the dynamics
of the certain class of non-Markovian systems whose 
simulation would
be otherwise too tedious because
of the requirement for large computational
resources.
\end{abstract}


\maketitle

\section{Introduction}
The description of the dynamics of open
quantum systems has attracted increasing
attention during the last few years \cite{Breuer02a}. The major
reason for this is the identification of
the phenomena of decoherence and dissipation, which
characterize the dynamics of open
quantum systems interacting with their
surroundings \cite{Joos03a}, 
as a main obstacle to the realization of quantum computers
and other quantum devices \cite{Nielsen00a}. Secondly, recent
experiments on engineering of environments \cite{Myatt00a}
have paved the way to new proposals aimed at
creating entanglement and superpositions
of quantum states exploiting decoherence and
dissipation \cite{Poyatos96a,Plenio02a}.

A common approach to the dynamics of open
quantum systems consists in deriving a master
equation for the reduced density matrix
which describes the temporal behavior
of the open system.
The solution for the master equation can then
be searched by using analytical
or simulation methods, or the combination
of both.

This article concentrates
on the developing of new Monte Carlo simulation methods 
for non-Markovian open quantum systems.
The general feature of the Monte Carlo methods
is the generation of an ensemble of stochastic realizations of
the state vector trajectories. The density matrix
and the properties of the system of interest
are then consequently calculated as an 
appropriate average of the generated ensemble.

Some common variants of the Monte Carlo
methods for open systems
include the Monte Carlo wave-function (MCWF) method 
\cite{Dalibard92a,Molmer96a}, the
quantum state diffusion (QSD) \cite{Gisin92a,Diosi98a,Strunz99a},
and the non-Markovian
wave function (NMWF) formulation unravelling the master equation
in an extended Hilbert space \cite{Breuer02a,Breuer99a,Breuer04a}.
The MCWF method has been
very successfully used to model the laser cooling of atoms.
Actually, 3D laser cooling
has so far been described only by MCWF simulations 
\cite{Molmer95a}.
QSD in turn has been found to have
a close connection to the decoherent
histories approach to quantum mechanics 
\cite{Diosi95a},
and NMWF method has been recently applied
to study the dynamics of quantum
Brownian particles \cite{Maniscalco04a,Maniscalco04c}.
The various Monte Carlo methods and related topics have been reviewed
e.g. in Refs.~\cite{Molmer96a,Carmichael93a,Gardiner96a,Plenio98a}

In general, simulating open quantum systems is a challenging
task. It has been shown earlier that the methods mentioned above
can solve a wide variety of problems. Nevertheless, sometimes
there
arise situations in which the complexity of the studied system
or the parameter region under study makes the requirement
for the computer resources so large that the solution
may become impossible in practice, though not
in principle.
Thus, it is important to assess the already existing
methods from this point of view, and 
develop new variants
to improve their applicability. This is the key
point of this article.

Here, we address the Monte Carlo simulation methods
for the short time-evolution of non-Markovian systems which are weakly
coupled to their environments.
In this case, the dynamics of the system may exhibit rich
features, whereas the weak coupling may
lead to extremely small quantum jump probabilities,
the consequence being unpractically large requirement
for the size of the generated Monte Carlo ensemble.
To overcome this problem, we present below a method 
which in general allows to reduce the ensemble
size.

By studying the Hilbert space path integral 
for the propagator of a piecewise deterministic
process (PDP) \cite{Breuer02a}, we show that part of the expectation
value of an arbitrary operator $A$ as a function of time $t$,
$\langle A \rangle (t)$, has  scaling  properties which can be
exploited in Monte Carlo simulations to speed up
the generation of the ensemble,
in the optimal case by several orders of magnitude.
We derive a scaling equation,
from which the result
for $\langle A \rangle (t)$
can be calculated,
all the quantities in the equation
being easily obtainable from the scaled
Monte Carlo simulations.

We concentrate first on the Lindblad-type
non-Markovian case which can be solved by the standard
MCWF method, and then focus
on the non-Lindblad-type case which
requires the use of the NMWF simulations in 
the doubled Hilbert space.

The paper is structured as follows.
Section \ref{sec:dyn} introduces the master equation,
the corresponding stochastic Schr\"{o}dinger equation, and
the appropriate simulation schemes
for the Lindblad- and non-Lindblad-type systems.
The Hilbert space path integral method is then used
to calculate the expectation value of an arbitrary
operator setting
the scene for the scaling method which 
is presented in Sec.~\ref{sec:scaling}. Section \ref{sec:examples}
shows explicitly how the scaling can be implemented
and demonstrates the usability of the method, 
for the example of quantum Brownian motion.
Finally Sec.~\ref{sec:conclusions} presents discussion
and conclusions.
 
\section{Dynamics of non-Markovian systems}\label{sec:dyn}

We describe first in Sec.~\ref{subsec:lindblad} the master equation for the 
Lindblad-type systems and the corresponding
standard MCWF method.
We then continue in Sec.\ref{subsec:nonlindblad} with the description
of the non-Lindblad-type case with the corresponding
stochastic Schr\"{o}dinger equation and NMWF unravelling in 
the doubled Hilbert space. The last subsection 
\ref{subsec:path}
presents the calculation of the expectation value
of an arbitrary operator $A$ with the Hilbert space path integral
method which paves the way for the scaling procedure. 

We begin by considering
master equations obtained from the
time-convolutionless projection operator technique
(TCL)
of the form  \cite{Breuer02a,Breuer99a}
\begin{eqnarray}
\frac{\partial}{\partial t} \rho\left( t \right) &=& A \left( t
\right)  \rho \left( t \right) +  \rho \left( t \right)
B^{\dag}\left( t \right)
\nonumber \\ 
&& +
 \sum_i C_i\left( t \right) \rho \left( t \right) D^{\dag}_i\left( t \right),
\label{eq:genmaster}
\end{eqnarray}
with time-dependent linear operators $A\left( t \right)$, $B\left(
t \right)$, $C_i\left( t \right)$, and $D_i\left( t \right)$.

\subsection{Lindblad-type case: master equation and MCWF method}
\label{subsec:lindblad}
A specific case of the master equation (\ref{eq:genmaster}) is the one of
Lindblad-type \cite{Gorini76a,Lindblad76a,Maniscalco04b}
\begin{eqnarray}
\frac{d}{dt}\rho(t) &=& -i \left[H_S,\rho(t)\right] + \sum_i \gamma_i(t)
\bigg\{ L_i \rho(t) L_i^\dagger -
\nonumber \\
&&
\left. 
 \frac{1}{2} L_i^\dagger L_i\rho(t)
- \frac{1}{2} \rho(t) L_i^\dagger L_i \right\},
\label{eq:master}
\end{eqnarray}
where $H_S$ is the system Hamiltonian, $\gamma_i(t)$
the time dependent decay rate to channel $i$, and $L_i$ is the corresponding
Lindblad operator. 

We define this non-Markovian master equation to be of Lindblad-type when
the time dependent decay coefficients $\gamma_i(t)\geq 0$ for
all times $t$,  and non-Lindblad type when $\gamma_i(t)$ acquire
temporarily negative values during the time-evolution \cite{Maniscalco04b}.
The Lindblad-type case can be treated with the standard 
MCWF method introduced in this subsection \cite{Dalibard92a}, 
and the non-Lindblad case
with the  NMWF method in the doubled Hilbert space
presented in the following subsection \cite{Breuer99a,Breuer02a}.

The core idea of the standard MCWF method is to generate
an ensemble of realizations for the state vector $\psi(t)$
by solving the
time dependent Schr\"{o}dinger equation
\begin{equation}
     i \hbar \frac{\partial \psi(t)}{\partial t}=
     H(t)\psi(t), 
\label{eq:Schrodinger}
\end{equation}
with the non-Hermitian Hamiltonian $H(t)$
\begin{equation}
     H(t)=H_{S}(t)+H_{DEC}(t) \label{eq:H},
\end{equation}
where $H_S(t)$ is the reduced system's Hamiltonian
and the non-Hermitian part $H_{DEC}(t)$ includes the sum over the various allowed decay
channels $i$,
\begin{equation}
     H_{DEC}(t)=-\frac{i\hbar}{2}\sum_{i}\gamma_i(t)L_i^{\dagger}L_i, \label{eq:HDec}
\end{equation}
where the jump operator $L_i$ for channel $i$
coincides with the Lindblad operator
appearing in the master equation (\ref{eq:master}). 

During a discrete time evolution step
of length $\delta t$ the norm of the state vector
may shrink due
to $H_{DEC}$. The amount of shrinking gives the probability of a
quantum jump to
occur during the short interval $\delta t$. Based on a random number one then
decides whether a quantum jump occurred or not. Before the next time
step is taken,
the  state vector of the system is renormalized. If and when a
jump occurs,
one performs a rearrangement of the state vector components
according to the jump
operator $L_i$, before renormalization of
$\psi$.

The jump probability corresponding to the decay
channel $i$ for each of the time-evolution steps $\delta t$ is 
\begin{equation}
     P_i(t)=\delta t \gamma_i(t)\langle\psi|L_i^{\dagger}L_i|\psi\rangle. \label{eq:Jp}
\end{equation}
The expectation value of an arbitrary operator A is then 
the ensemble average over the generated realizations
\begin{equation}
\langle A \rangle (t) = \frac{1}{N}\sum_{i=1}^N \langle \psi_i | A | \psi_i \rangle,
\end{equation}
where $N$ is the number of realizations.

\subsection{Non-Lindblad-type case:
Stochastic Schr\"odinger equation and NMWF method in the doubled Hilbert space}
\label{subsec:nonlindblad}

The solution of the general master equation (\ref{eq:genmaster})
can be obtained 
by using the
NMWF unravelling in the doubled Hilbert space
$\widetilde{\cal H}={\cal H}_S\oplus{\cal
H}_S$ where ${\cal H}_S$ is the Hilbert space of the 
system \cite{Breuer02a,Breuer99a} .
The state of the system is described by a pair of
stochastic state vectors
\begin{equation}
\theta\left( t \right) = \left(\begin{array}{c}
      \phi  \left( t \right) \\
      \psi\left( t \right)
\end{array} \right),
\end{equation}
such that $\theta(t)$ becomes a stochastic process in the doubled
Hilbert space $\widetilde{\cal H}$.
Denoting the corresponding probability density functional
by $\widetilde{P}[\theta,t]$, we can define the reduced
density matrix as 
\begin{equation}
\rho(t)=\int D \theta D \theta^{\ast} |\phi\rangle \langle\psi |
\tilde{P}[\theta,t].
\label{eq:rhobasic}
\end{equation}

The time-evolution of $\theta\left( t \right)$ can be described as
a piecewise deterministic process (PDP) 
and the corresponding stochastic Schr\"{o}dinger equation
reads \cite{Breuer02a}
\begin{eqnarray}
&&
d\theta(t) = -i G\left(\theta,t\right)dt +
\nonumber \\
&& 
\sum_i \left\{ \frac{\|\theta(t)\|}{\| J_i(t)\theta(t)\|}J_i(t)\theta(t)-
\theta(t)\right\} dN_i(t),
\label{eq:SSE}
\end{eqnarray}
where the Poisson increments satisfy the equations
\begin{eqnarray}
dN_i(t)~dN_j(t) &=& \delta_{ij}dN_i(t), \nonumber \\
E\left[dN_i(t)\right] &=& \frac{\|J_i(t)\theta(t)\|^2}{\|\theta(t)\|^2}dt,
\end{eqnarray}
and the non-linear operator $G(\theta,t)$ is defined as
\begin{equation}
G(\theta,t) = \left[ F\left( t \right)+\frac{1}{2}\sum_i \frac{\| J_i\left( t
\right)\theta\left( t \right) \|^2}{\| \theta\left( t \right)
\|^2} \right]
\theta\left( t \right),
\end{equation}
with the time-dependent operators
\begin{eqnarray}
F\left( t \right) &=& \left(
\begin{array}{cc}
      A\left( t \right) & 0 \\
       0 &  B\left( t \right)
\end{array} \right)
\nonumber \\
J_i\left( t \right) &=&  \left( \begin{array}{cc}
      C_i\left( t \right) & 0 \\
      0                    & D_i\left( t \right)
\end{array} \right),
\end{eqnarray}
where $A\left( t \right)$, $B\left( t \right)$, $C_i\left( t
\right)$, and $D_i\left( t \right)$ are the operators appearing in
Eq.~(\ref{eq:genmaster}).

The deterministic part of the PDP is obtained by solving the
following differential equation
\begin{eqnarray}
i\frac{\partial}{\partial t} \theta \left( t \right) =
G(\theta,t),
\end{eqnarray}
and the jumps of the PDP take the form
\begin{equation}
\theta \left( t \right) \rightarrow \frac{\| \theta \left( t
\right) \|} {\| J_i\left( t \right)\theta \left( t \right) \|}
\left(
\begin{array}{c}
      C_i\left( t \right) \phi \left( t \right) \\
      D_i\left( t \right) \psi \left( t \right)
\end{array} \right).
\end{equation}
Once the ensemble of stochastic realizations
has been generated, one can then calculate the density matrix
of the reduced system from Eq.~(\ref{eq:rhobasic}).

\subsection{The Hilbert space path integral for the propagator of the PDP
and the expectation value of arbitrary operators}
\label{subsec:path}

For simplicity, we present here the Hilbert space path integral for 
the Lindblad-type case. The derivation of the non-Lindblad-type case follows
closely the presentation below.

We assume that the initial state of the system
is a pure state $\psi_0$.
In this case the propagator $T$ of the PDP (conditional transition probability) 
coincides with the probability density functional $P$ of the stochastic
process \cite{Breuer02a}
\begin{equation}
P\left[\psi, t\right]=
T\left[\psi,t | \psi_0, t_0 \right].
\end{equation}
This quantity describes the probability of the system being
in the state $\psi$ at time $t$ when
it was in the state $\psi_0$ at some earlier time $t_0$.
For short time non-Markovian evolutions and weak couplings, we assume 
that the maximum 
number of jumps per realization is one. Thus, the expansion of the propagator
$T$ in terms of number of jumps contains two terms:
deterministic evolution without jumps $T^{(0)}$ and paths
with one jump $T^{(1)}$
\begin{eqnarray}
T\left[\psi,t | \psi_0, 0 \right] =
T^{(0)}\left[\psi,t | \psi_0, 0 \right] +
T^{(1)}\left[\psi,t | \psi_0, 0 \right].
\label{eq:T}
\end{eqnarray}

With the assumptions above, the expectation value
of an arbitrary operator $A$ at time $t$ can be calculated as
\cite{Breuer02a}
\begin{eqnarray}
\langle A \rangle \left( t \right) &=&  
\int D \psi D \psi^* \langle\psi | A | \psi \rangle 
T\left[\psi,t | \psi_0, t_0 \right] 
\nonumber \\
&=& 
\int D \psi D \psi^* \langle\psi | A | \psi \rangle 
\nonumber \\
&&
\left\{T^{(0)}\left[\psi,t | \psi_0, t_0 \right]
+ T^{(1)}\left[\psi,t | \psi_0, t_0 \right]\right\}. 
\label{eq:P} 
\end{eqnarray}
By calculating  $T^{(0)}$ and $T^{(1)}$, see Appendix \ref{app:a}, we obtain
for $\langle A \rangle (t)$
\begin{eqnarray}
\langle A \rangle \left( t \right) &=&
\int D \psi D \psi^* \langle\psi | A | \psi \rangle 
\nonumber \\
&&
\times \left\{
\left[ 1-\int_0^t ds \sum_i \gamma_i(s) 
\| L_i  g_s(\psi_0) \|^2 \right]
\right. \nonumber \\
&&
\times\delta\left( \psi - g_t\left(\psi_0\right) \right)
\nonumber \\
&&
+
\int_0^t ds \int D \psi_1 D \psi_1^* 
\int D \psi_2 D \psi_2^*
\nonumber \\
&&
\times\delta\left( \psi-g_t\left( \psi_2  \right)  \right)
\sum_i \gamma_i(s) \| L_i \psi_1  \|^2
\nonumber \\
&&
\times\delta\left( \frac{L_i \psi_1}{\| L_i\psi_1 \|} - \psi_2   \right)
\delta\left( \psi_1 - g_s\left( \psi_0  \right)  \right)
\Bigg\}. 
\label{eq:P2} 
\end{eqnarray}

Here, terms of the form $\delta\left( \psi - g_t\left(\psi_0\right) \right)$
are the functional delta-functions and 
the deterministic evolution of $\psi_0$
 according to the non-Hermitian Hamiltonian $H$ is
given by
\begin{equation}
g_t\left(\psi_0\right) = 
\frac{\exp\left( -i\int_0^{t}H(t')dt'\right) \psi_0}
{\|\exp\left( -i\int_0^{t}H(t')dt'\right) \psi_0\|}.
\end{equation} 
The physical interpretation of Eq.~(\ref{eq:P2}) is straightforward.
The expectation value of $A$ is calculated with respect
to all possible paths of $\psi$ with appropriate  weights.
The first term in the curly brackets is 
the no-jump evolution of $\psi$ multiplied with the
corresponding probability of no-jumps.
The second term includes the integration over
all possible jump times and jump routes 
with the appropriate transition rates for the one
jump realization. 

\section{Scaling}\label{sec:scaling}

Denoting the expectation value of $A$ with respect
to the no-jump evolution 
as
\begin{equation}
\langle A \rangle_0 (t) = 
\langle \psi = g_t\left( \psi_0  \right) | A |
 \psi = g_t\left( \psi_0  \right) \rangle,
\label{eq:a0}
\end{equation}
we obtain from Eq.~(\ref{eq:P2})
\begin{eqnarray}
&&
\langle A \rangle \left( t \right) - 
\langle A \rangle_0 (t) 
=
 \int D \psi D \psi^*
\langle \psi | A | \psi \rangle 
\nonumber \\
&& 
\times
\left\{ -  
\delta \left(  \psi - g_t(\psi_0)  \right) \int_0^t ds \sum_i  \gamma_i(s)
\|L_i  g_s(\psi) \|^2
\nonumber \right. \\
&&+
\int_0^t ds \int D \psi_1 D \psi_1^* \int D \psi_2 D \psi_2^*~
\delta \left(  \psi - g_t(\psi_2)  \right) 
\nonumber \\
&&
\times
\sum_i  \gamma_i(s) \| L_i \psi_1  \|^2
\delta \left(  \frac{L_i \psi_1}{\| L_i\psi_1 \|}  -\psi_2\right)
\nonumber \\
&&
\times
\delta \left(  \psi_1 - g_s(\psi_0)  \right)
\bigg\}.\label{eq:aini}
\end{eqnarray}
This equation leads to the first key observation of the paper.
We notice that $\langle A \rangle \left( t \right) - 
\langle A \rangle_0 (t)$ (but not $\langle A \rangle \left( t \right)$
alone) is directly proportional 
to the transition rates of the type
\begin{equation}
W \left[ \psi_2 | \psi_1 \right] = 
\sum_i \gamma_i(t)  \| L_i \psi_1 \|^2
\delta \left[ \frac{L_i \psi_1}{\| L_i\psi_1 \|} -
\psi_2 \right].\label{eq:W}
\end{equation}
In the corresponding Monte Carlo simulations for the case we
are considering, the required size 
of the generated ensemble is related to the transition rates $W$
since the rate defines the number of jumps. In more detail, if the
total (cumulative) jump probability for the time evolution
period of interest is $P_c$, we need on average to
generate $1/P_c$ realizations to produce one realization
which has a jump. To achieve good statistical accuracy
we need obviously
a large enough number of jumps and
the minimum condition for the required
ensemble size $N$ becomes $N\gg 1/P_c$.

This leads us to the following observation which can be used
to optimize the ensemble size of the Monte Carlo simulations
(within the approximations we use).
We can artificially increase the number of jumps
by scaling up the transition rate $W$ by a factor of $\beta$.
At the same time we must leave the non-Hermitian
Hamiltonian $H$ unscaled since the ensemble
average contribution given by realizations
with $H$ only (no jumps) appears
on the l.h.s. of the equation (\ref{eq:aini}).
In other words, we are not allowed to
scale the deterministic evolution of the state
vector (which includes also the rotation of
the state vector towards the state with the smallest decay
rate $\gamma_i$) but only increase the number of jumps
by scaling up the transition rates by a factor of $\beta$.
In the simulation this can be done easily
multiplying the jump probabilities for various
decay channels by a same factor $\beta$.
An explicit example how to do this for both
of the cases we are considering, Lindblad- and
non-Lindblad-type, is shown in the next section.

The question is now how we can calculate
from the scaled simulations the result we are looking for,
namely the expectation value for arbitrary
operator $A$ as a function of time $\langle A \rangle (t)$.
It can be shown, see Appendix \ref{app:b},
that the final result for $\langle A \rangle (t)$ starting from
Eq.~(\ref{eq:aini}) can be obtained as
\begin{eqnarray}
\langle A \rangle (t) &=& 
\left( 1-\frac{P_{tot}(t)}{\beta} -
\frac{1}{\beta}\frac{N-N_j(t)}{N} \right) \langle A \rangle_0 (t) 
\nonumber \\
 &&
+\frac{1}{\beta} \bar{\langle A \rangle}_{tot}(t).
\label{eq:fina}
\end{eqnarray}
This equation is the main result of the paper.
It shows that the ensemble average of the scaled simulations
can be used
to calculate the result for the original problem
we are interested in.
In this equation,
$P_{tot}(t)$ is the total transition rate
(see Appendix \ref{app:b}),
$N$ is the size of the ensemble, $N_j(t)$ the number
of jumps in the simulations
as a function of time, $\beta$ the scaling factor,
$\langle A \rangle_0(t)$ the expectation value
with respect the deterministic time evolution 
(see Eq.~(\ref{eq:a0})), and $\bar{\langle A \rangle}_{tot}(t)$
the ensemble average from the modified
simulations 
where the scaling has been used (see the discussion above).
All of the quantities on the r.h.s. can be easily
calculated in the simulation. Actually, 
from a technical point of view,
the only difference
between the scaled and unscaled simulations 
is that in the former one we have
to keep track of the number of jumps
as a function of time. A task which
can be easily done in the simulations.
We also note that
at time $t=0$, $P_{tot}(0)=0$, $N_j(0)=0$, 
$\bar{\langle A \rangle}_{tot}(0) = \langle A \rangle_0(0)$
and we obtain correctly for time $t=0$:
$\langle A \rangle (0)=\langle A \rangle_0(0)$.

Thus, we can optimize the ensemble size by using
the following procedure in the Monte Carlo simulations:
i)  Scale up jump probabilities by suitable factor $\beta$.
ii) Leave decay rates
$\gamma_i(t)$ untouched in the non-Hermitian Hamiltonian $H$
 iii) Calculate the result
for $\langle A \rangle (t)$ from Eq.~(\ref{eq:fina}).

It is worth to emphasize here a common feature of 
Monte Carlo wave-function simulations.
The deterministic evolution caused by 
the non-Hermitian Hamiltonian $H$ changes the relative
weights of the occupied states due to the different
decay rates of the various states. The scaling procedure
incorporates this rotation by adding to the scaled
ensemble average result [the second term on the r.h.s. of Eq.~(\ref{eq:fina})]
contribution from the deterministic
evolution calculated with the appropriate weight (the first term).

Since the reduction in the required
ensemble size is directly proportional
to the used scaling factor $\beta$,
the issue is now how large scaling factor 
we can use to optimize the simulations. 
The scaling method that we have developed
is valid when
there is maximally
one jump per realization.
 This condition has to 
hold also for the scaled simulations as well.
As soon
as the scaling factor is so large that 
realizations with two or more jumps begin
to occur, additional error (with respect
to the normal statistical error of Monte Carlo
simulations) starts to appear. 
In other words, the probability
of having two jumps per realization
has to be much smaller than the one jump probability.
If the total probability for
one jump is $P_c$ (see the discussion above),
the probability for two jumps equals $P_c^2$
and the estimate for additional 
error is simply given by $P_c^2/P_c=P_c$.
Thus we can use the scaling
factor which increases the jump
probabilities e.g. to the order of $0.01$
introducing a manageable $1\%$ error
in addition to the normal statistical error
of the Monte Carlo simulations.

For the standard Monte Carlo simulations
there exists a corresponding measurement scheme
interpretation based on the continuous monitoring of the environment
of the system. 
The scaling technique modifies the Monte Carlo
simulation method in such a way that
the measurement scheme interpretation
is lost.

The scaled simulations correspond to a stochastic
Schr\"odinger equation where the deterministic
part generated by $G$, see Eq.~(\ref{eq:SSE}), remains the same but the jump
part is scaled with $\beta$, i.e. the expectation value
of the Poisson increment becomes 
\begin{equation}
E\left[dN_i(t)\right] = \beta \gamma_i(t)\|L_i\psi\|^2dt.
\end{equation}
Thus the stochastic Schr\"odinger equation does not have 
a corresponding
master equation, and actually does not
need to have one for the scaling to work. 
This is because we are not looking
for two master equations whose
results are scalable from each other.
Rather the key point is 
to modify in a suitable way the
equations for the simulations in order
to make them faster and more efficient.

Summarizing, we have demonstrated above how the scaling works
for the Lindblad-type master equation with
time-dependent but always positive decay coefficients
$\gamma_i(t)$. For this the standard Monte Carlo wave-function
method can be used \cite{Dalibard92a,Plenio98a}. 
In a similar way, it can be shown
that the scaling works also for the non-Lindblad-type
case where $\gamma_i(t)$ may acquire temporarily negative
values. In this case one needs to use the doubled Hilbert
space unravelling \cite{Breuer02a}. We show examples
of the scaling for both of these cases in the next Section.

\section{Examples for scaling}\label{sec:examples}

The discussion above shows how it is possible to reduce the size
of the generated ensemble in the Monte Carlo simulations
for non-Markovian systems. It is worth
noting that for the Markovian case the scaling
is not needed because the jump probabilities
can be increased trivially by increasing e.g. the time
step size $\delta t$ in the simulations. For the non-Markovian
case this does not work because the main features
of the open system dynamics may be given by the 
time dependence of the decay rates, and
$\delta t$ has to be kept small compared
to the temporal variations of the decay coefficients.

We show below two examples for the scaling. In these examples
we use the scaling factors $\beta=10^4$ and $10^5$ while
the generated ensembles have the sizes of the order of $10^5$.
In other words, without the scaling, the solution of the presented
problems would require at least $10^9$ ensemble members. 

To demonstrate the scaling, we  perform
the simulations for the short time non-Markovian dynamics
of a quantum Brownian particle (damped
harmonic oscillator) \cite{Maniscalco04a,Maniscalco04c}.
We demonstrate both
the Lindblad-type, and non-Lindblad type cases. 

The dynamics of a harmonic oscillator linearly coupled to a
quantized reservoir, modelled as an infinite chain of quantum
harmonic oscillators, is described, in the secular approximation, by
means of the following generalized master equation
\cite{Intravaia03a,Maniscalco04b}
\begin{eqnarray}
&&\frac{ d \rho(t)}{d t}= \frac{\Delta(t) \!+\! \Gamma (t)}{2}
\left[2 a \rho(t) a^{\dag}- a^{\dag} a \rho(t)  - \rho(t)
a^{\dag} a \right]
\nonumber \\
&& +\frac{\Delta(t) \!-\! \Gamma (t)}{2} \left[2 a^{\dag} \rho(t)
a - a a^{\dag} \rho(t) - \rho(t) a a^{\dag}
 \right]. \nonumber \\
 \label{eq:mqbm}
\end{eqnarray}
In the previous equation, $a$ and $a^{\dag}$ are the annihilation
and creation operators, and $ \rho(t)$ the density matrix of the
system harmonic oscillator. The time dependent coefficients
$\Delta(t)$ and $\Gamma(t)$ appearing in the master equation are
known as diffusion and dissipation coefficients, respectively
\cite{Intravaia03a,Maniscalco04a}. 

For an Ohmic reservoir spectral
density with Lorentz-Drude cut-off,
the expression for 
$\Delta(t)$ is \cite{Caldeira83a,Maniscalco04a}
\begin{eqnarray}
\Delta(t) &=& 2 \alpha^2 k T \frac{r^2}{1+r^2} \left\{ 1
- e^{-\omega_c t} \left[ \cos (\omega_0 t)\right.\right.
 \nonumber \\
&&
-  (1/r)  \left. \sin (\omega_0 t )\right] \big\},
\label{eq:deltaHT}
\end{eqnarray}
where the assumption of the high temperature
reservoir has been used.
The dissipation coefficient $\Gamma(t)$ can be written 
\begin{equation}
\Gamma (t)\! \!=\!\! \frac{\alpha^2 \omega_0 r^2}{r^2+1} \Big[1
\!-\! e^{- \omega_c t} \cos(\omega_0 t) \! - r e^{- \omega_c t}
\sin( \omega_0 t )  \Big]. \label{gammasecord}
\end{equation}
Here, $r=\omega_c/\omega_0$ is the ratio between the environment
cut-off frequency $\omega_c$ and the oscillator frequency
$\omega_0$, $\alpha$ is the dimensionless coupling constant,  $k$
the Boltzmann constant, and $T$ the temperature. 
When $r>1$, the decay coefficients $\Delta(t)\pm\Gamma(t)>0$
for all times,
and the master equation is of Lindblad type.   
When $r<1$, the decay coefficients $\Delta(t)\pm\Gamma(t)$
acquire temporarily negative values
and the master equation is of non-Lindblad type \cite{Maniscalco04a}.
For Lindblad-type master equation, one can apply the standard 
MCWF method (Sec.~\ref{sec:Lin} ) where as the non-Lindblad-type case requires
the application of the NMWF method
in the doubled Hilbert space (Sec.~\ref{sec:NonLin})
 
To demonstrate that the scaling works (in addition 
to the rigorous proof presented above), we compare below
the results obtained from the simulations to the
exact analytical results \cite{Maniscalco04a,Intravaia03b}. 
  
\subsection{Lindblad-type master equation and MCWF simulations}\label{sec:Lin}

For the Lindblad type case we choose parameters
$2 \alpha^2 k T /\omega_c=1.2\times10^{-6}$,  $r=10$, 
$\alpha^2 \omega_0 /\omega_c=0.5\times10^{-8}$, 
and the scaling
factor $\beta=10^4$.
The initial state of the system is chosen
to be a coherent state
$|\xi=\sqrt{2}\rangle$ such that, at $t=0$, 
$\langle n \rangle = |\xi|^2=2$.
 We emphasize that the present 
paper generalizes the scaling method we have used
in \cite{Maniscalco04a} for
initial Fock states to 
 arbitrary system
Hamiltonians and arbitrary initial states. 

The non-Hermitian part of the Hamiltonian
is now given by [see Eqs.~(\ref{eq:master}), (\ref{eq:H}), and (\ref{eq:mqbm})]
\begin{equation}
     H_{DEC}=-\frac{i\hbar}{2}\left\{
 [\Delta(t)-\Gamma(t)]aa^{\dagger} +
[\Delta(t)+\Gamma(t)]a^{\dagger}a
\right\}.
\end{equation}
The jump probabilities for each
time step $\delta t$ and decay channel $i$
are now modified so that the jump probability
for channel $1$ (jump up, absorption 
of one quantum of energy from the environment) is
\begin{equation}
     P_1(t)=\beta~\delta t [\Delta(t)-\Gamma(t)]
\langle\psi|aa^{\dagger}|\psi\rangle,
\end{equation}
and for channel $2$ (jump down,
emission of one quantum of energy into the environment)
\begin{equation}
     P_2(t)=\beta~\delta t [\Delta(t)+\Gamma(t)]
\langle\psi|a^{\dagger}a|\psi\rangle. 
\end{equation}
The ensemble average is then calculated
in the usual Monte Carlo way, as presented
in Section \ref{subsec:lindblad} and the
simulation results plugged into Eq.~(\ref{eq:fina}) 
to get the final result.

\begin{figure}[tb]%[ht!]
\centering
\includegraphics[scale=0.4]{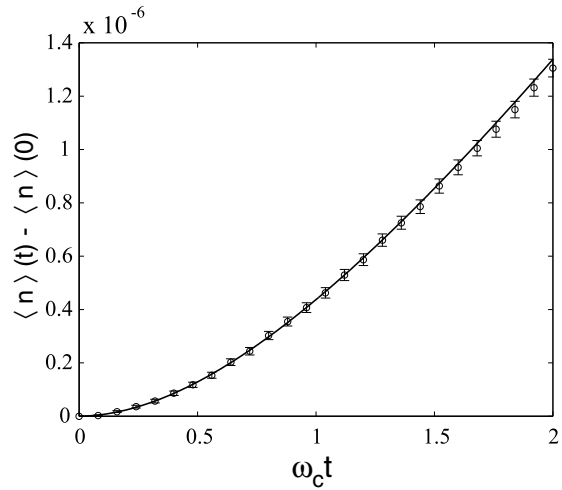}
\caption[f2]{\label{fig:eg1}
Comparison between analytical (solid line) and scaled simulation
results (circles) with the bars of the standard error
for the Lindblad-type case. The figure shows
the behavior of the expectation value of the quantum number
$\langle n\rangle$ as a function of time. The initial state
of the system is a coherent state $|\xi=\sqrt{2}\rangle$.
For the parameters used here,
the scaling reduces the required ensemble size by a factor 
on the order of $10^{4}$.
The simulation here contains $6\times 10^{5}$ realizations.
}
\end{figure}

Figure \ref{fig:eg1} shows the excellent match between 
the analytical curve and the simulations using
the scaling. For the discussion of the  analytical solution, see
Ref.~\cite{Maniscalco04a}. The results confirm once more the validity of the scaling
procedure and show the short time quadratic non-Markovian behavior of the 
average quantum number $\langle n \rangle=\langle a^{\dag}a\rangle$
of the oscillator.
Moreover, for the parameters used here,
the scaling reduces the required ensemble size by a factor of $10^{4}$.
The simulation here contains $6\times 10^{5}$ realizations. 

%%%% fig %%%%%
%% end fig %%%

\subsection{Non-Lindblad-type unravelling in the doubled Hilbert space}\label{sec:NonLin}

For the non-Lindblad-type case we choose
the following parameters $2 \alpha^2 k T /\omega_c=2.4\times10^{-6}$, $r=0.1$, 
$\alpha^2 \omega_0 /\omega_c=0.5\times10^{-8}$, and 
the scaling factor $\beta=10^5$.
As initial state we choose a
superposition of Fock states $\psi=(|0\rangle+|1\rangle)/\sqrt{2}$.

The doubled Hilbert space state vector for the harmonic oscillator
reads
\begin{equation}
\theta(t) = \left(
 \begin{array}{c}
      \phi  \left( t \right) \\
      \psi\left( t \right)
\end{array} \right)
= \left(
\begin{array}{c}
\sum_{n=0}^{\infty} \phi_n(t) |n\rangle \\
 \sum_{n=0}^{\infty} \psi_n(t) |n\rangle \\
\end{array} \right),
\end{equation}
where  $\phi_n(t)$ and $\psi_n(t)$ are the probability amplitudes
in the Fock state basis.

By comparing Eq.~(\ref{eq:mqbm}) with the master equation
(\ref{eq:genmaster}), the operators $A(t)$ and $B(t)$
have to be chosen as
\begin{eqnarray}
A(t) &=& B(t)= -i \omega_0 a^{\dag} a - \frac{1}{2} \left\{
\left[\Delta(t)+\Gamma(t)\right]a^{\dag} a+ \right.
\nonumber \\
&& \left. \left[\Delta(t)-\Gamma(t)\right]aa^{\dag} \right\}.
\end{eqnarray}
Accordingly, the operators $C_i$ and $D_i$ are
\begin{eqnarray}
C_1(t)&=&D_1(t)= \sqrt {|\Delta(t)-\Gamma(t)|} a^{\dag},
\nonumber \\
C_2(t)&=&D_2(t)= \sqrt{|\Delta(t)+\Gamma(t)|} a
\end{eqnarray}
and the corresponding operators $J_i$,
become
\begin{eqnarray}
J_1(t)= \sqrt {|\Delta(t)-\Gamma(t)|} \left(
\begin{array}{cc}
 {\rm sgn}\left[\Delta(t)-\Gamma(t)\right] a^{\dag}& 0 \\
0& a^{\dag}\\
\end{array} \right)
\nonumber \\
J_2(t)= \sqrt {|\Delta(t)+\Gamma(t)|} \left(
\begin{array}{cc}
 {\rm sgn}\left[\Delta(t)+\Gamma(t)\right] a& 0 \\
0& a\\
\end{array} \right).
\end{eqnarray}

%%%% fig %%%%%
\begin{figure}[tb]%[ht!]
\centering
\includegraphics[scale=0.4]{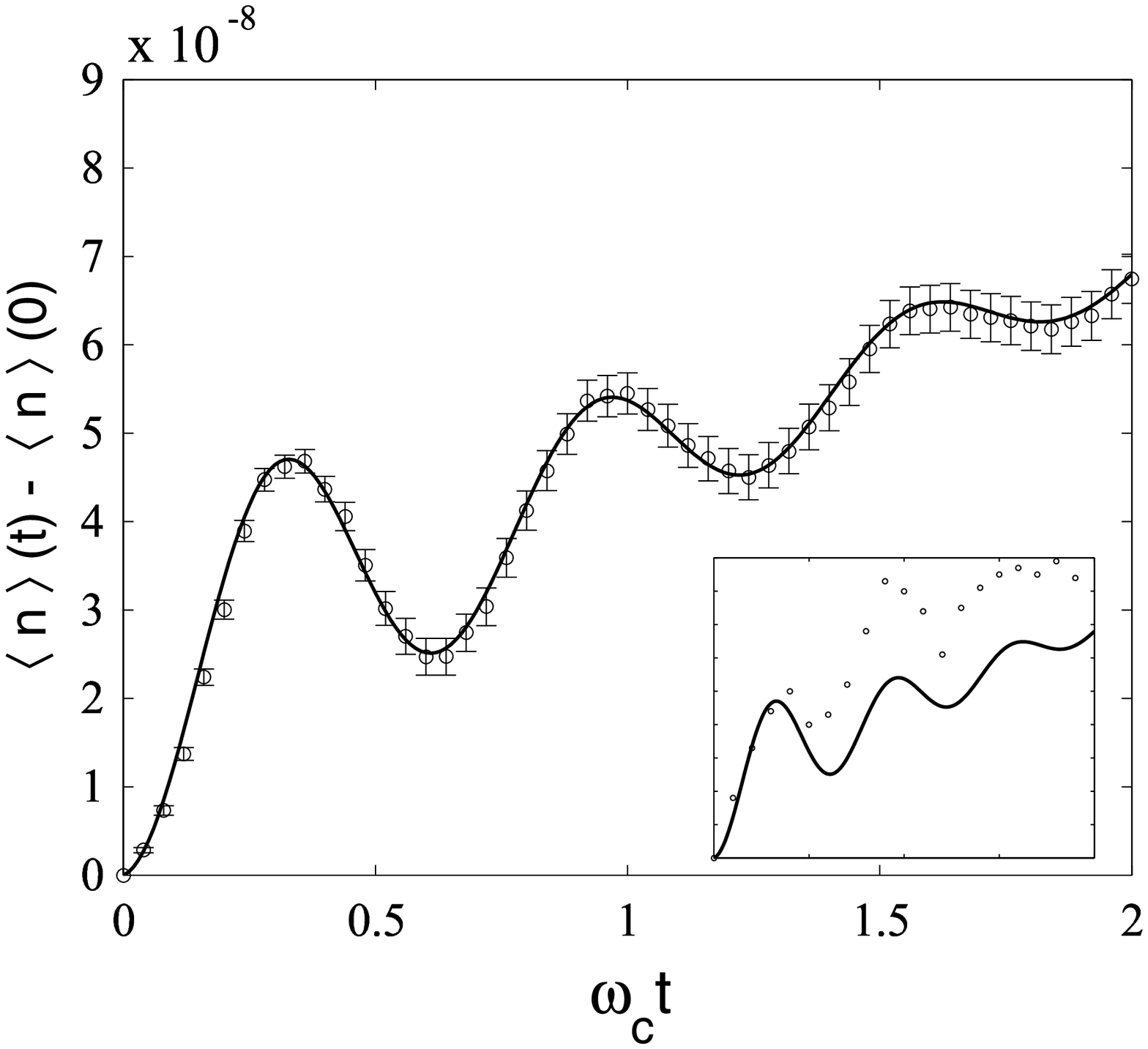}
\caption[f2]{\label{fig:eg2}
Comparison between analytical (solid line) and scaled simulation
results (circles) with the bars of the standard error for the non-Lindblad-type case. 
The figure shows
the behavior of the expectation value of the quantum number
$\langle n\rangle$ as a function of time. The initial state
of the system is the superposition of Fock states
$(|0\rangle+|1\rangle)/\sqrt{2}$.
For the parameters used here,
the scaling reduces the required ensemble size atleast by a factor of $10^{4}$.
The simulation contains $6\times10^{5}$ realizations.
The inset shows (in the same scale as the main plot)
the poor match between the analytical result (solid line) and 
the simulation result without the scaling (circles) with
$6\times10^{8}$ realizations which is three orders of magnitude larger
than used in the scaling (see text).
}
\end{figure}
%% end fig %%%

The statistics of the quantum jumps
is described by the waiting
time distribution function $F_w(\tau)$ which represents the
probability that the next jump occurs within the time interval
$[t,t+\tau)$. $F_w(\tau)$, derived from the properties of the
stochastic process, reads
\begin{equation}
F_w(\tau)=1-\exp\left[-\int_0^{\tau} \sum_{i=1,2}
P_i\left(s\right)ds\right],\label{eq:fw}
\end{equation}
where for channel 1 (jump up, the system absorbs a quantum of
energy from the environment)
\begin{equation}
P_1(t)=\beta~\frac{|\Delta(t)-\Gamma(t)|}{\| \theta \left( t \right)
\|^2} \sum_{n=0}^{\infty} (n+1)\left[ |\phi_n(t)|^2  +
|\psi_n(t)|^2\right], \label{p1}
\end{equation}
and for channel $2$  (jump down, the system emits a quantum of energy
into the environment)
\begin{equation}
P_2(t)=\beta~\frac{|\Delta(t)+\Gamma(t)|}{\| \theta \left( t \right)
\|^2} \sum_{n=0}^{\infty} n \left[ |\phi_n(t)|^2  +
|\psi_n(t)|^2\right]. \label{p2}
\end{equation}
Here, the probabilities are scaled with a factor of $\beta$
according to the scaling scheme presented above. 
When the jump occurs, the choice of the decay channel is made
according to the factors $P_1(t)$ and $P_2(t)$. The times at which
the jumps occur are obtained from Eq. (\ref{eq:fw}) by using the
method of inversion \cite{Breuer02a}.

Figure~\ref{fig:eg2} displays the short time 
oscillatory non-Markovian behavior of the average
quantum number $\langle n \rangle$.
This type of behavior is
studied in detail in Ref.~\cite{Maniscalco04a}.
The results show the excellent match between 
the exact analytical solution and the simulation results using
the scaling with $6\times10^{5}$ realizations.
Again, the results confirm the validity of the scaling
procedure. Moreover, the inset shows a very poor match
between the non-scaled simulations with $6\times10^{8}$
realizations~\cite{Piilo05a} and justifies the claim that
the reduction in the ensemble size is at least on the order
of $10^4$ when the scaling procedure is used.

The reduction of the ensemble size can
be estimated also by calculating the maximum
jump probability of a single realization. 
In the example considered here, the maximum probability
is of the order of $10^{-7}$, in other words on average
an ensemble size of $10^7$ produces one jump event
in the unscaled simulations.
We estimate that one needs several hundreds jumps
in the simulations to produce accurately the rich dynamical 
features of the heating function displayed in Fig.~\ref{fig:eg2},
and consequently the requirement for the ensemble size
is at least $10^9$ without the scaling. Thus, the reduction
in the ensemble size by the scaling method is again found to be at least
on the order of $10^4$.

\begin{figure}[tb]%[ht!]
\centering
\includegraphics[scale=0.4]{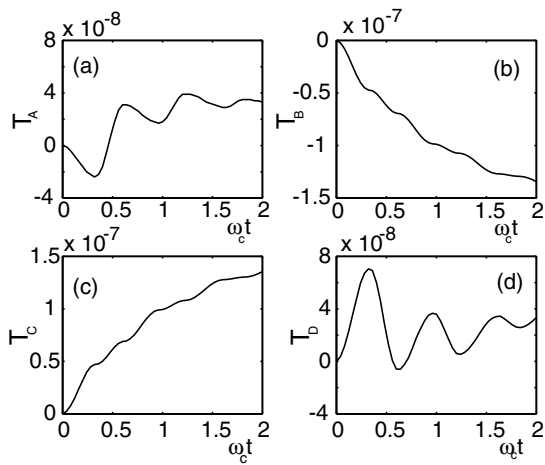}
\caption[f2]{\label{fig:terms}
Contribution from the various terms of the scaling equation
(\ref{eq:fina}).
(a) $T_A = \langle A\rangle_0(t)$, 
(b) $T_B = -P_{tot}(t)\langle A \rangle_0(t) / \beta$, 
(c) $T_C = -[(N-N_j(t))/(\beta N)] \langle A\rangle_0(t)$,
(d) $T_D = \langle {\bar A}\rangle _{tot}(t)/\beta$.
The terms has been shifted to start from the same
initial value for easier comparison.
Here $A$ is the number operator $A = a^{\dag}a$. 
The final result presented in Fig.~2. is mostly given as a sum
of the terms
displayed in (a) and (d).
}
\end{figure}

It is interesting to compare the various terms
in the scaling equation (\ref{eq:fina}) in the non-Lindblad
case. Figure \ref{fig:terms} shows the four
terms of the scaling equation (\ref{eq:fina}).
One can notice that two of the terms
practically cancel each other and the final
result is mostly given by the two
terms presented in Figs.~\ref{fig:terms} (a) and (d).

\section{Discussion and conclusions} \label{sec:conclusions}

We have demonstrated 
a scaling method for Monte Carlo wave-function
simulations 
which can reduce the size of the generated 
ensemble
by several orders of magnitude
especially for weakly coupled
non-Markovian systems.
The scaling is based on the notion that
once in the simulations the jump probabilities are scaled,
and the deterministic evolution given by the
non-Hermitian Hamiltonian left untouched,
one can obtain the time evolution
of the observables of interest
from the
scaling equation (\ref{eq:fina}).

The scaling has been used in a restricted form,
for a specific physical system,
in Ref.~\cite{Maniscalco04a}.
In that case the initial state of the system was
a Fock state. Here, we present a generalized
scaling scheme which is able to treat arbitrary
initial states of the system and arbitrary
Hamiltonians.
We emphasize that the scaling method works very well
for solving the short time dynamics of non-Markovian,
systems, which
bear importance e.g. for the decoherence
studies for quantum information processing \cite{Alicki02a}.

%%%% fig %%%%%
%% end fig %%%

In general, non-Markovian systems, even when they are weakly
coupled to their environments, can posses rich
dynamical features despite of the fact that the quantum jump
probability per stochastic realization is small 
during the time evolution period of interest (see the examples above).
This is the key area where the scaling method we have presented
is useful. The small jump probabilities due to the weak coupling
can lead to the situations where the requirement
for the size of the generated ensemble in the Monte Carlo
wave function simulations is unconveniently large. 
In these cases, the scaling method can be used to reduce
and optimize the generated ensemble size for
efficient numerical simulation of weakly coupled
non-Markovian systems. 

The scaling method presented here can be used
when the master equation of the open quantum system
can be expressed in the general form of Eq.~(\ref{eq:genmaster})
obtained by the time-convolutionless projection operator
techniques (the one-jump restriction still applies,
see below). To compare our method
to the other simulation methods for non-Markovian systems 
one should actually compare the validity of the TCL
with respect to the methods presented e.g.~in Refs.~\cite{Imamoglu96a,Yu99a,Gambetta02a}.
Thus, making a rigorous
comparison is an involved
task and is left for future studies. 
We initially note here that our method
is not restricted with respect to the temperature
of the environment (while method presented in
Ref.~\cite{Gambetta02a} is valid for the zero-temperature bath)
and is valid, at least in principle,
to the order used in the TCL expansion
of master equation to be unravelled
(while method presented in Ref.~\cite{Yu99a} is post-Markovian,
i.e. first order correction to Markovian dynamics).
However, it is worth mentioning that the validity
of the TCL expansion is crucially related to the
existence of the TCL generator (see e.g.~page $447$
of Ref.~\cite{Breuer02a}).

The scaling method is limited to the cases
where there is maximally one jump per realization
in the generated Monte Carlo ensemble.
Moreover, it is also important to note that the
same restriction applies also for the scaled simulations. 
These limits can be easily checked by calculating
the jump probabilities  from Eqs.~(\ref{eq:Jp}) and (\ref{eq:fw})
for the time period of interest
or by monitoring the number of jumps in the simulations.
As soon as more than one jump per realization
in the scaled simulations begin to occur, one can
estimate the error by calculating the ratio between
the two-jump and the one-jump probabilities
per realization.
In the examples we have described, we have not
used very aggressive optimization of the ensemble
size (though the ensemble size reduction is on the order of $10^4$),
and no error has been introduced. This has been confirmed
by monitoring the jumps in the simulations: no two-jump
realizations was generated. Thus, the error bars displayed
in the Figs. (\ref{fig:eg1}) and (\ref{fig:eg2}) correspond to the usual
statistical error (standard deviation) of the Monte Carlo ensemble. 

In conclusion, the scaling method has limitations
(one jump per realization) 
but it is interesting to note that in the region
where the method
can not be applied (more than one jump
per realization), it is not needed. This 
is because in this region there already occurs large enough
number of jumps enhancing the statistical accuracy
of the simulations. In other words, the problem which
the scaling solves appears only
within the region of validity of the method.

\acknowledgments
The authors thank H.-P. Breuer for discussions in Freiburg
and acknowledge CSC - the Finnish IT center for science -
for the computer resources.
This work has been financially supported by 
the Academy of Finland (JP, project no. 204777), the Magnus
Ehrnrooth Foundation (JP), and the Angelo Della Riccia Foundation
(SM).

\appendix

\section{Hilbert space path integral} \label{app:a}

Expanding the exponential waiting time distribution $F$ and
taking into account the terms corresponding
to maximum one jump per realization for short times
and weak couplings,
the contribution to the propagator from the path without
the jumps is \cite{Breuer02a}
\begin{eqnarray}
&&
T^{(0)}\left[\psi,t | \psi_0, 0 \right] =
\left( 1-F\left[ \psi_0,t \right] \right)
\delta \left[ \psi-g_t\left(\psi_0\right) \right] =
\nonumber \\
&&
\left( 1-\int_0^t ds \sum_i \gamma_i(s) 
\| L_i  g_s(\psi_0) \|^2 \right) \times
\nonumber \\
&&
\delta\left( \psi - g_t\left(\psi_0\right) \right)
\label{eq:T0}
\end{eqnarray}
where 
$\delta\left( \psi - g_t\left(\psi_0\right) \right)$
is the functional delta-function and 
the deterministic evolution according to the 
non-Hermitian Hamiltonian $H$ is
given by
\begin{equation}
g_t\left(\psi_0\right) = 
\frac{\exp\left( -i\int_0^{t}H(t')dt'\right) \psi_0}
{\|\exp\left( -i\int_0^{t}H(t')dt'\right) \psi_0\|}.
\end{equation} 
where
\begin{equation}
H = H_S -\frac{\imath \hbar}{2} \sum_i \gamma_i (t)L_i^{\dag} L_i.
\label{eq:nonHH}
\end{equation}

By using the recursion relation for the 
propagator $T$ \cite{Breuer02a}
and neglecting
the terms of the  order
of $\gamma_i(t)^2$ or higher, one can now calculate
the contribution of the one jump path to the propagator
as
\begin{eqnarray}
&&
T^{(1)}\left[\psi,t | \psi_0, 0 \right] =
\int_0^t ds \int D \psi_1 D \psi_1^* 
\int D \psi_2 D \psi_2^*~
\nonumber \\
&&
\delta\left( \psi-g_t\left( \psi_2  \right)  \right)
\sum_i \gamma_i(s) \| L_i \psi_1  \|^2
\delta\left( \frac{L_i \psi_1}{\| L_i\psi_1 \|} - \psi_2   \right)
\nonumber \\
&&
\delta\left( \psi_1 - g_s\left( \psi_0  \right)  \right).
\label{eq:T1}
\end{eqnarray}
where the transition rate summed over the decay channels is
\begin{equation}
W \left[ \psi_2 | \psi_1 \right] = 
\sum_i \gamma_i(s)  \| L_i \psi_1 \|^2
\delta \left[ \frac{L_i \psi_1}{\| L_i\psi_1 \|} -
\psi_2 \right],
\end{equation}
The physical interpretation of Eq.~(\ref{eq:T1}) is straightforward.
The integrations sums over the various one jump routes and
over all the possible jump times.

\section{Expectation value}\label{app:b}

In the simulations we scale up the jump probabilities
by a factor $\beta$, and leave the non-Hermitian Hamiltonian
as it is [includes also $\gamma_i(t)$], we get
the corresponding equation for Eq.~(\ref{eq:aini}) as 
\begin{eqnarray}
&&
\beta\left[\langle A \rangle \left( t \right) - 
\langle A \rangle_0 (t) \right]
=
 \int D \psi D \psi^*~ 
\langle \psi | A | \psi \rangle
\nonumber \\
&& 
\times
\left\{ -  
\delta \left(  \psi - g_t(\psi_0)  \right) 
\int_0^t ds \sum_i \beta \gamma_i(s)
\|L_i  g_s(\psi) \|^2
\nonumber \right. \\
&&
+
\int_0^t ds \int D \psi_1 D \psi_1^* \int D \psi_2 D \psi_2^*~
\delta \left(  \psi - g_t(\psi_2)  \right) 
\nonumber \\
&&
\times
\sum_i \beta \gamma_i(s) \| L_i \psi_1  \|^2
\delta \left(  \frac{L_i \psi_1}{\| L_i\psi_1 \|}  -\psi_2\right)
\nonumber \\ 
&&
\times
\delta \left(  \psi_1 - g_s(\psi_0)  \right)
\bigg\}.\label{eq:newsc}
\end{eqnarray}
For scaling to work,
we have to be able to extract from the simulations
the information on the r.h.s. of this equation.

This can be done as follows.
We note the first term on the r.h.s.
of Eq.~(\ref{eq:newsc}) as 
\begin{equation}
\bar{\langle A \rangle_0} (t) = 
P_{tot}(t) \langle A \rangle_0 (t) 
\end{equation}
where $P_{tot}(t)$ is the total transition rate
\begin{equation}
P_{tot}(t)=\int_0^t ds \sum_i \beta \gamma_i(s)
\|L_i  g_s(\psi) \|^2.
\end{equation}

Furthermore, we denote by $\bar{\langle A \rangle_1}$ the second term on the r.h.s.
of Eq.~(\ref{eq:newsc}) as
\begin{eqnarray}
&&
\bar{\langle A \rangle_1} (t) =
\int D \psi D \psi^*~  
\langle \psi | A | \psi \rangle
\nonumber \\
&&
\left[   
\int_0^t ds \int D \psi_1 D \psi_1^* \int D \psi_2 D \psi_2^*~
\delta \left(  \psi - g_t(\psi_2)  \right) 
\right.\nonumber \\
&&
\left.
\times\sum_i \beta \gamma_i(s) \| L_i \psi_1  \|^2
\delta \left(  \frac{L_i \psi_1}{\| L_i\psi_1 \|}  -\psi_2\right)
\delta \left(  \psi_1 - g_s(\psi_0)  \right)
\right] 
\nonumber \\
&=&
\frac{N_j(t)}{N}\sum_{i=1}^{N_j(t)} \langle \psi_i(t) | A |\psi_i(t) \rangle /N_j(t),
\end{eqnarray}
where $N_j(t)$ is number of jumps, and $N$ the total number of realizations.
Here, the second part is the one jump contribution to
the expectation value, expressed formally, and the summation is carried over those
realizations that have jumped till time $t$. 
The corresponding simulation
presentation (simulation average) is given in the last part.

Now the ensemble average of all realizations $\bar{\langle A \rangle}_{tot}(t)$,
the quantity which we can easily calculate in the simulation,
is given by
as a sum of 0 and 1 jump realization contributions
\begin{equation}
\bar{\langle A \rangle}_{tot}(t)
=
\frac{N-N_j(t)}{N}\langle A \rangle_{0}(t) + 
\bar{\langle A \rangle}_{1}(t).
\label{eq:atot}
\end{equation}

Equation (\ref{eq:newsc}), which includes the quantity we are interested in,
can now be written as
\begin{eqnarray}
&&
\beta\left[\langle A \rangle \left( t \right) - 
\langle A \rangle_0 (t) \right] = -
\bar{\langle A \rangle}_{0}(t) +
\bar{\langle A \rangle}_{1}(t) = 
\nonumber \\
&&
-
P_{tot}\langle A \rangle_0 (t) -
\frac{N-N_j}{N}\langle A \rangle_{0}(t)+
\bar{\langle A \rangle}_{tot}(t).
\end{eqnarray}

From this equation we easily obtain the final result for the expectation
value of arbitrary operator $A$ in compact form as
\begin{eqnarray}
\langle A \rangle (t) &=& 
\left( 1-\frac{P_{tot}(t)}{\beta} -
\frac{1}{\beta}\frac{N-N_j(t)}{N} \right) \langle A \rangle_0 (t)
\nonumber \\
&& +
\frac{1}{\beta} \bar{\langle A \rangle}_{tot}(t).
\end{eqnarray}


\begin{thebibliography}{99}
\bibitem{Piilo05a}
This is the largest ensemble size we can conveniently generate
with the current computer resources at our disposal.
\end{thebibliography}
\end{document}